\pdfoutput=1

\documentclass[aps,prl,onecolumn]{revtex4} 

\usepackage{color, datetime}
\usepackage{graphicx}
\usepackage{amsmath, amsthm, amssymb}
\usepackage{pdfpages}
\def\({\left(}
\def\){\right)}
\def\[{\left[}
\def\]{\right]}
\DeclareMathOperator\sign{sgn}

\begin{document}
\title{Modulation of mechanical resonance by chemical potential oscillation in graphene}

\author{Changyao Chen$^{1\dagger}$, Vikram V. Deshpande$^2$, Mikito Koshino$^3$, Sunwoo Lee$^4$, Alexander Gondarenko$^1$, Allan H. MacDonald$^5$, Philip Kim$^6$ \& James Hone$^{1\ast}$}


\affiliation{
${^1}$Department of Mechanical Engineering, Columbia University, New York, NY 10027, USA\\
${^2}$Department of Physics and Astronomy, University of Utah, Salt Lake City, UT, 84112, USA\\
${^3}$Department of Physics, Tohoku University, Sendai 980-8578, Japan\\
${^4}$Department of Electrical Engineering, Columbia University, New York, NY 10027, USA\\
${^5}$Department of Physics, University of Texas, Austin, TX 78712, USA\\
${^6}$Department of Physics, Harvard University, Cambridge, MA, 02138, USA\\
{$^\dagger$ Current address: Center for Nanoscale Materials, Argonne National Laboratory, Lemont, IL, 60439, USA}\\
\normalsize{$^\ast$ Corresponding email: jh2228@columbia.edu}
}

\begin{abstract}
The classical picture of the force on a capacitor assumes a large density of electronic states, such that the electrochemical potential of charges added to the capacitor is given by the external electrostatic potential and the capacitance is determined purely by geometry.  Here we consider capacitively driven motion of a nano-mechanical resonator with a low density of states, in which these assumptions can break down.  We find three leading-order corrections to the classical picture:  the first of is a modulation in the static force due to variation in the internal chemical potential; the second and third are change in static force and dynamic spring constant due to the rate of change of chemical potential, expressed as the quantum (density of states) capacitance.  As a demonstration, we study a capacitively driven graphene mechanical resonators, where the chemical potential is modulated independently of the gate voltage using an applied magnetic field to manipulate the energy of electrons residing in discrete Landau levels.  In these devices, we observe large periodic frequency shifts consistent with the three corrections to the classical picture.  In devices with extremely low strain and disorder, the first correction term dominates and the resonant frequency closely follows the chemical potential. The theoretical model fits the data with only one adjustable parameter representing disorder-broadening of the Landau levels. The underlying electromechanical coupling mechanism is not limited the particular choice of material, geometry, or mechanism for variation in chemical potential, and can thus be extended to other low-dimensional systems.
\end{abstract}
\maketitle

The calculation of the force between plates of a parallel-plate capacitor by energy methods is a classic problem that illustrates the importance of correct definition of the system's free energy:  considering only the energy stored in the capacitor incorrectly predicts a repulsive force of $\frac 1 2 \(\text d C / \text d z\)\phi^2$, where $C$ is the capacitance, $\text d C / \text d z$ is its spatial derivative, and $\phi$ is the electrostatic potential difference. The correct result (an attractive force $-\frac 1 2 \(\text d C / \text d z\)\phi^2$) is only obtained when the work done by the battery to maintain constant voltage is taken into account\cite{Feynman_lecture}. This result assumes a simple system of metallic capacitors with large density of states (DOS), such that the chemical potential $\mu$ is constant and the electrostatic potential is identical to the voltage applied by the battery.  This assumption breaks down in nanoscale systems, in which the DOS can be much smaller and $\mu$ is no longer fixed, an effect widely studied in electronic devices\cite{Klitzing_prl_1980,Fulton_prl_1987,Luryi_apl_1988,Park_nature_2002,Ilani_nphys_2006,Xia_nnano_2009}. However, the role of finite DOS in the force between capacitor plates has not been rigorously investigated. Understanding the corrections to the classical model is of fundamental interest and important for modeling of nano-mechanical systems in the atomically thin limit\cite{Steele_science_2009,Lassagne_science_2009}.  Here we examine the case of a resonant nano-mechanical device, and show that the first-order correction to the resonant frequency consists of three terms, one proportional to $\mu$ and two proportional to its derivative $\text d \mu / \text d n$, with the latter represented through the inverse quantum capacitance $C_Q^{-1} = \frac 1{Ae^2} \frac{\text d \mu}{\text d n}$ ($A$ is the sample size and $e$ is the electron charge).  The phenomenon is explored experimentally by examining the behavior of a graphene resonator, in which $\mu$ is tuned by a perpendicular magnetic field $B$ at \emph{fixed electrochemical potential}.  We observe strong frequency shifts that are periodic in 1/$B$, and can be quantitatively described by the theoretical model.  We further find that, in the limit of an ultra-clean device with low tension, the first term can provide the dominant tuning effect, demonstrating that a nano-mechanical device can be used to track its own chemical potential.

We model a generic mechanical resonator as a mass on a nonlinear spring, capacitively actuated by a nearby gate electrode, as shown in fig. 1. A direct current (DC) voltage $V_g$ applied to the gate can both increase the resonant frequency by exerting a static force to change the equilibrium deflection $z_e$,  and decrease the frequency due to the nonlinear electrostatic potential, an effect known as electrostatic spring softening\cite{Lifshitz_2009}. The two effects provide an effective spring constant given by:
\begin{equation}
\label{nonlinear_spring}
k_\text{eff} = k + \eta \alpha z_e^2 - \frac 1 2 \frac {\text d^2 C_\text{total}}{\text d z^2} V_g^2, 
\end{equation}
where $k$ is the linear spring constant, $\alpha$ is the elastic nonlinear coefficient, and $\eta$ is the geometric coefficient that depends on the exact form of static deflection (see Supplementary Information, section 2). In rigid microelectromechanical devices (MEMS), the last capacitive softening term provides the dominant tuning effect, whereas in atomically thin materials under low tension, the second mechanical stiffening term can dominate.

The role of the finite density of states is modeled by considering the variation in $\mu$ with carrier density $n$, such that the electrostatic potential is given by $\phi=V_g - \mu/e$, and by including $C_Q$ in series with $C_g$. In this case, we first find that the static capacitive force is given by (Supplementary Information, section 3):
\begin{equation}
\label{Delta_k1}
F = - \frac 12 \(V_g - \frac {\mu}{e}\)^2\(1 - 2\frac{C_g}{C_Q} \)\frac{\text d C_g}{\text d z}.
\end{equation}
When $V_g$ is kept constant and $\mu$ and $C_Q$ are modulated, the change in static force is approximated as:
\begin{equation}
\label{Delta_k2}
\Delta F \approx - \[ \frac {\Delta \mu} e  + C_g V_g \Delta \( \frac1 {C_Q} \)\] V_g \frac{\text d C_g}{\text d z}. 
\end{equation}
The change in static force shifts the equilibrium deflection $z_e$, thus the resonant frequency, and alters the curvature of the potential energy at the equilibrium position, as determined from the spatial derivative of $\Delta F$.  The leading term in the latter is given by:
\begin{equation}
\Delta k = \frac {\text d \Delta F}{\text d z} =   \( V_g \frac{\text d C_g}{\text d z} \)^2 \Delta  \(\frac{1}{C_Q}\).
\end{equation}
Combining the effects from eq. \eqref{Delta_k1} and \eqref{Delta_k2}, we arrive at the total frequency shift as:
\begin{eqnarray}
\label{freq_shift}
\Delta f_\text{total} &=& - \Re_V \( \frac{\Delta \mu}{e}\) 
              -  \Re_V C_g V_g \Delta \( \frac{1}{C_Q}\)\
              + \frac {f_0} {2k_0}  \( V_g \frac{\text d C_g}{\text d z} \)^2 \Delta  \(\frac{1}{C_Q}\),
\end{eqnarray}
where $\Re_V = |\text d f_0 / \text d V_g|$ indicates how easily the resonant frequency $f_0$ can be tuned by electrostatic potential, $k_0$ is the unmodulated spring constant, $\Delta \mu$ is change in chemical potential, and $\Delta\( {1}/{C_Q}\)$ represents the change in inverse quantum capacitance. The three terms in eq. \eqref{freq_shift} are referred to below as $\Delta f_1$, $\Delta f_2$, and $\Delta f_3$, respectively.  Importantly $\Delta f_1$ is a direct measure of variation in the chemical potential, whereas $\Delta f_2$ and $\Delta f_3$ depend on the change in quantum capacitance. We also note that these terms are closely related to experiments with nanotube quantum dot mechanical resonators\cite{Steele_science_2009,Lassagne_science_2009}, in which $\Delta f_{2,3}$ are analogous to the diverging charge susceptibility of the quantum dot that produces dips in the nanotube mechanical frequency at charge transitions, and $\Delta f_1$ plays a similar role that changes the static tension and the mechanical frequency from discretely charging the quantum dot.

We examine this phenomenon experimentally using graphene mechanical resonators\cite{Bunch_Science_2007,chen_nnano_2009}, where a transverse magnetic field $B$ is used to tune  $\mu$ and $C_Q$ independently of $V_g$. The samples are in a three-terminal configuration (fig. 2a) with a local gate (LG) placed few hundred nanometers away from the suspended graphene. The graphene is assumed to have built-in strain $\varepsilon_0$ from the fabrication process, with additional strain induced by the DC gate voltage $V_g$. We use a continuum mechanical model for calculation of $f(V_g)$ at different levels of $\varepsilon_0$ (fig. 2b, also Supplementary Information, section 3). Fig. 2c shows $\Re_V$ as a function of $\varepsilon_0$, at three different values of $V_g$, for a 2 by 2 $\mu$m single layer graphene resonator. Thus, in devices with low $\varepsilon_0$, $\Re_V$ can be large and $\Delta f_{1,2}$ are maximized whereas in devices with large $\varepsilon_0$, $\Re_V$ is substantially smaller and $\Delta f_{1,2}$ should be minimized relative to $\Delta f_3$.  We note that previous studies of graphene resonators in high magnetic fields\cite{Singh_apl_2012} utilized samples with high tension and therefore observed only the effects of $\Delta f_3$; in the experimental work below, we focus on samples with ultralow tension to directly observe the effects of chemical potential variation.

In a transverse magnetic field, electronic states of graphene form discrete Landau levels (LLs), and $\mu$ oscillates in a sawtooth pattern with increasing $B$. The oscillation in $\mu$ is reflected in the longitudinal electrical resistivity\cite{Zhang_nature_2005, Novoselov_nature_2005,Neto_rmp_2009} (Shubnikov-de Haas oscillations), and in the magnetization $M=-\text d \mu / \text d B$ (de Haas-van Alphen oscillations, typically detected by torque magnetometry\cite{Eisenstein_prl_1985}), but can also be studied explicitly using a sensitive single-electron transistor electrometer\cite{Zhitenev_nature_2000,Feldman_science_2012}.  In a single particle picture, the dependence of $\mu$ and $C_Q$ on $B$ is determined by the filling fraction $\nu = 2\pi n\hbar/eB$ ($\hbar$ is the reduced Planck's constant), and disorder.  We construct a simple model\cite{Koshino_prb_2007} consisting of up to 20 Gaussian-broadened LLs with disorder-induced width $\Gamma$ (in units of $v_F\sqrt{2e\hbar B}$, where $v_F$ = 10$^6$ m/s is the Fermi velocity). The DOS is given by $D(E) = \text d n / \text d \mu \propto \sum\limits^N \exp[-(E-E_N)^2/\Gamma^2]$, where $E_N$ is center of the $N^\text{th}$ LL. Within each LL, $E_N$ of single-layer graphene evolves with $B$ as $E_N(B)=v_F \sign(N) \sqrt{2e\hbar B |N|}$. Here $\mu$ is assigned as the highest filled energy level, and $C_Q$ is defined as $Ae^2 D(E)$, where $A$ is the sample area. Figures 2d, 2e and 2f show the simulated $\Delta \mu$ as a function of $B$ and accompanying frequency shifts calculated from each of the terms in eq. \eqref{freq_shift} for three different combinations of high/low tension and disorder.  In the case of low tension but large disorder (fig. 2d), $\Delta f_{1,2,3}$ are all reflected in the total frequency shift, but LL broadening largely obscures the variation in $\mu$. In the case of high tension and small disorder (fig. 2e), we found that $\Delta f_3 > \Delta f_{1,2}$ as expected, leading to sharp spikes in frequency at LL transitions.   Finally, in the case of both low tension and low disorder, the frequency shift closely (but not completely) follows the contribution of $\Delta f_1$ and the sawtooth variation of $\mu$, with spikes at the transitions between LLs due to $\Delta f_{2,3}$.

This simulation motivates the use of graphene samples with very low built-in strain and disorder, in order to directly observe variation in $\mu$.  To prevent contamination, we prepare samples by direct exfoliation of graphite over pre-patterned electrodes\cite{xu_apl_2010}, and further clean the graphene by Joule heating in vacuum at low temperature\cite{Moser_apl_2007,Bolotin_ssc_2008}. The electronic quality is examined through separate measurements (Supplementary Information, section 1), and can yield charged impurity density as low as $\sim 8 \times 10^9$ cm$^{-2}$. For mechanical resonance measurements, the graphene is actuated electrostatically by adding a small radiofrequency drive voltage to the gate, and the gate-drain current at the same frequency is read out by a vector network analyzer; on resonance, the changing capacitance due to mechanical motion causes a measurable peak in the current\cite{xu_apl_2010, Chen_phd_2013}. The absence of DC bias avoids Joule heating and possible Lorentz forces. We scan $V_g$ up to $\pm$10 V at zero magnetic field to determine $\Re_V$ and use the continuum mechanical model to extract $\varepsilon_0$.  The use of substrate-fixed electrodes eliminates tensioning due to thermal contraction of the metal\cite{chen_nnano_2009,Singh_nano_2010}, and we are able to achieve $\Re_V$ as large as 10 MHz/V, with $\varepsilon_0 \sim 10^{-4}$.  

Figure 3a shows the response of mechanical resonance to applied magnetic filed $B$, for device D1 with $\Re_V$ of 2 MHz/V. There is no obvious frequency shift with $B$, except for regimes between LLs, where sharp spikes are evident but the signal becomes weak. Such observations agree well with the predictions for small $\Re_V$ and experimental results reported by Singh $et~al.$\cite{Singh_apl_2012}. Fig. 3b shows the same measurements on device D2 (length 2.4 $\mu$m, width 3.2 $\mu$m), with $\Re_V$ of 10 MHz/V. The data reveals a repeated pattern of oscillations that are periodic in $1/B$ (Supplementary Information, section 4), allowing us to directly extract the carrier density $n$ of 2.4$\times$10$^{11}$ cm$^{-2}$.  We use the measured value of $n$ at 5 different values of $V_g$, combined with finite element simulation (COMSOL Multiphysics) to obtain the effective displacement $z_e$ and $\text d C_g / \text d z$ as $C_g/z_e$ as a function of $V_g$. Therefore, the only free parameter left to fit eq. \eqref{freq_shift} is the disorder level $\Gamma$, which will determine $\Delta \mu$ and $\Delta(1/C_Q)$. We found that a $\Gamma$ of 0.1 gives satisfactory result.  The corresponding $\mu (B)$, that gives rise to $\Delta f_1$,  is overlaid as a yellow curve in fig. 3b. For comparison, we also plot $E_N(B)$ for the first five LLs as dotted green lines. We found that $\mu$ traces $E_N$ closely in each LL, confirming that the frequency can closely track $\mu$ in high quality samples.  The detailed fit between the model and the data is shown in fig. 3c.  Here, the solid points represent the resonant frequency exctracted at each value of $B$ from the data in fig. 3b, and the line represents the prediction of eq. \eqref{freq_shift} with $\Gamma = 0.1$.  The fit is excellent and validates the model presented above.  

We briefly consider the region between LLs, where the frequency shift should be primarily sensitive to $C_Q$. At high magnetic fields, when the sample enters the well-developed quantum Hall regime, its bulk becomes insulating and the edges host dissipationless one-dimensional edge states. This will not change the static charge on the sample and the static force due to $V_g$, and therefore first two terms of eq. \eqref{freq_shift} should be unaffected.  However, when the bulk becomes sufficiently insulating such that its RC charging time is greater than the mechanical resonance period, the dynamic charging should be determined by the geometric capacitance between the gate and the edge channels, which can be an order of magnitude smaller than $C_g$  (see Supplementary Information, section 5).  This has two consequences. First, both the third term in eq. \eqref{freq_shift} and the classical electrostatic softening term in eq. \eqref{nonlinear_spring} should decrease with the decreasing dynamic capacitance; in general, the classical term will dominate and the frequency spike will be larger than predicted by eq. \eqref{freq_shift}.  Second, both the radiofrequency drive and capacitive displacement current will also decrease, leading to substantial loss of signal.   As a consequence, we can use the loss of signal to directly determine the onset of the bulk insulating behavior.  In fig. 3b,c (also see fig. S11), this is seen to occur for the final spike, at the transition between $N$ = 2 and $N$ = 1, indicated by the dashed line in fig. 3c.  For this transition, we expect that the model will not fit the data well, and indeed, the magnitude and width of the frequency spike, where observable, seem to exceed the prediction. Nevertheless, the best fit of the limited data in the region between LLs yields $C_Q$ with values similar to previous measurements of high quality graphene samples\cite{Martin_nphys_2009,Yu_pnas_2013}. More detailed modeling and alternative transduction techniques ($e.g.$ optical detection) may allow better study of the detailed behavior between LLs.   

Figure 4a shows similar data at different values of $V_g$, with fits using $\Gamma$ values within $\pm$ 10\%. Since we are able to ``read off'' chemical potential variation directly through mechanical resonance shifts, we can then track the chemical potential as a function of the filling fraction $\nu$. Fig. 4b shows $\Delta \mu (\nu)$ at four different values of $V_g$: there are distinct jumps of $\mu$ at $\nu$ = 4($N$ + 1/2), which represent the energy gaps between neighboring LLs. We linearly extrapolate to extract the energy gaps, as plotted in fig 4c.   In addition, the same analysis allows extrapolation to determine the  chemical potential at zero magnetic field $\mu_0 =  \hbar v_F \sqrt{\pi n}$.  As shown in fig. 4d, this follows $\sqrt{V_g}$ as expected. Therefore the mechanical measurements can, under appropriate conditions, provide a direct means to monitor the chemical potential evolution of the underlying system, in a straightforward manner. We expect to extend our measurements into extreme quantum regimes at larger $B$, where additional chemical potential jumps would suggest the formation of many-body incompressible states.

Our nanomechanical technique thus provides access to a key thermodynamic quantity (chemical potential) in a self-contained manner. Specifically, our system allows measurements of both the energy gaps, and the gradual energy transition between gaps, which would otherwise require specialized instrumentation\cite{Zhitenev_nature_2000,Feldman_science_2012}. In the future, measurements of other correlation effects like negative compressibility\cite{Efros_ssc_1988,Kravchenko_prb_1990}, fractional quantum Hall effects\cite{bolotin_nature_2009,du_nature_2009,Dean_nphys_2011,Feldman_science_2012,Dean_nature_2013,Ponomarenko_nature_2013} and Wigner crystallization\cite{PhysRevB.34.2681}, all of which have thermodynamic signatures, should be possible using this technique. Finally, we emphasize that the electromechanical mechanism identified in the present work is not specific to graphene. Indeed, any atomically thin mechanical resonator with a quasi-2D electronic structure will demonstrate this sensitive response to external fields via an electrostatically controlled mechanism.

The authors thank Nigel Cooper, Igor Aleiner, Brian Skinner, Gary Steele for helpful discussions; David Heinz and Andrea Young for help in building the measurement setup; Noah Clay for fabrication support; Kin-Chung Fong, Tony Heinz, Andrea Young and Arend van der Zande for helpful comments. P.K. and J.H. acknowledge Air Force Office of Scientific Research Grant No. MURI FA955009-1-0705. P.K. acknowledges FENA.

[Author Contributions] C.C. and V.V.D. fabricated and characterized the samples, developed the measurement technique and performed the experiments, data analysis and theoretical modeling. M.K. performed disorder-broadened calculations. S.L. and A. G. helped in sample fabrication. A. H. M. provided theoretical support. P.K. and J.H. oversaw the project. C.C., V.V.D., P.K. and J.H. co-wrote the paper. All authors discussed and commented on the manuscript.

[Author Information] The authors declare no competing financial interests. Correspondence and requests for materials should be addressed to J.H. (email: jh2228@columbia.edu).

\pagebreak

\subsection{Device Fabrication}

High resistivity silicon wafers ($>$20,000 $\Omega$ cm) are used in the experiments described in the main text, in order to minimize the parasitic capacitance in the radio frequency (RF) range. 290 nm of thermal oxide is then grown to have best optical contrast for identifying single layer graphene in later studies.

Source and drain electrodes (1 nm Cr, 15 nm Au) are first patterned by DUV photo-lithography (ASML 300C DUV stepper, Cornell Nanofabrication Facility). Next, a dry etch (CF$_{4}$) creates 200 nm deep trenches between source and drain electrodes, and finally local gate electrodes (1 nm Cr, 20 nm Au) are deposited in the trenches with careful alignment. The detail of the fabrication is described elsewhere\cite{Chen_phd_2013}. 

Graphene samples are prepared by mechanical exfoliation of Kish graphite. The color contrast of the substrate-supported region aids in identification of thin flakes crossing the electrodes. The precise number of layers is subsequently confirmed by Raman spectroscopy.

\subsection{Device Characterization}

Suspended graphene samples are electrically tested at room temperature under vacuum before cool-down. Only samples that possess reasonable transconductance ($\Delta R > 100 \Omega$ with in 10 V change of gate voltage) are chosen for further investigation.

For low temperature measurements in a magnetic field, samples are wire-bonded and loaded into a home-built insert for the Quantum Design Physical Properties Measurement System (PPMS, 1.8 K base temperature) with 8 semi-rigid coaxial cables to transmit both DC and RF signals. Most suspended graphene samples show improved quality (narrower Dirac peak) upon cooling. Sample quality is further improved by gentle current annealing\cite{Bolotin_ssc_2008}. After such treatments, the smallest full-width at half-maximum (FWHM) of the Dirac peak is about 0.2 V, corresponding to a disorder density of $\sim 8 \times 10^{9}$ cm$^{-2}$. To avoid collapsing the graphene due to electrostatic force, we only apply gate voltage within $\pm$10 V. 





\newpage
\begin{figure}
\includegraphics[width=1\textwidth]{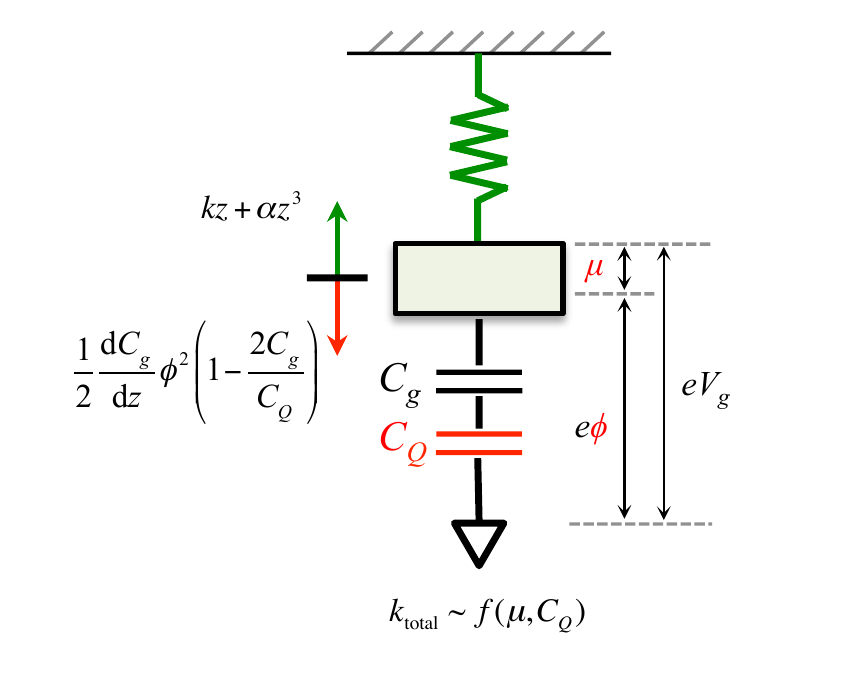}
\caption{Mass on a nonlinear spring balanced with electrostatic force. The force exerted by the spring varies
nonlinearly with the displacement $z$, and the electrostatic force is determined not only by gate
capacitance $C_g$, electrostatic potential $\phi$, but also quantum capacitance $C_Q$. For fixed
electrochemical potential $V_g$, $\phi$ is directly modified by chemical potential $\mu$, therefore the
total spring constant $k_{\text{total}}$ is modulated by both $\mu$ and $C_Q$, and magnified by the
nonlinear effect (large $\alpha$).}
\end{figure}

\begin{figure}
\includegraphics[width=1\textwidth]{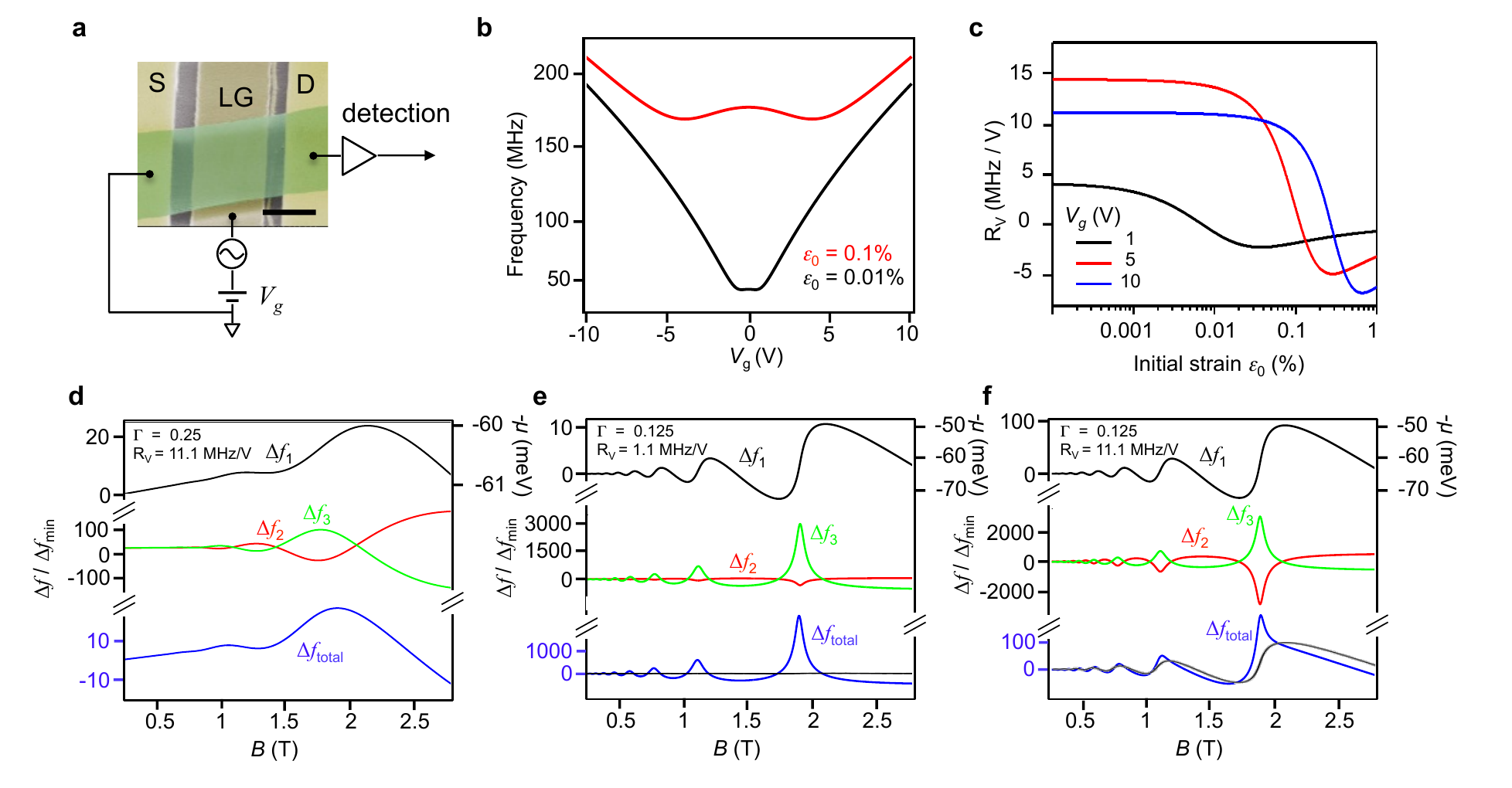}
\caption{Mechanical resonance modulation. \textbf{(a)} Schematic of the graphene mechanical resonator with source (S), drain (D) and local gate (LG) contacts. Scale bar: 1 $\mu$m. \textbf{(b)} Simulated resonant frequency for a 2 by 2 $\mu$m graphene resonator with 120 nm gap, at different initial strain $\varepsilon_0$. \textbf{(c)} Calculated $\Re_V$ as function of initial strain $\varepsilon_0$ for single layer graphene resonator, at different gate voltage $V_g$. The negative $\Re_V$ indicates electrostatic softening effect. \textbf{(d)}, \textbf{(e)}, \textbf{(f)} Simulated frequency shifts under high disorder and large $\Re_V$, \textbf{(d)}; low disorder and small $\Re_V$ \textbf{(e)}; low disorder and large $\Re_V$ \textbf{(f)}. Top panel: $\mu$ (to the right) and accompanied frequency oscillation $\Delta f_1$ (to the left) as function of magnetic field $B$. Middle panel: corresponding frequency shifts $\Delta f_2$ and $\Delta f_3$ (see main text). Bottom panel: the total frequency shift as function of $B$, with $\Delta f_1$ overlaid on top (black), except for \textbf{(d)}. For all the simulations, $V_g$ = 10V, $\varepsilon_0$ = 0.01\%, $Q$ = 2000, signal-to-noise ratio = 10 dB, $f_0$ = 161.4 MHz, and $f_{\text{min}}$ is 360 Hz assuming 100 Hz measurement bandwidth.}
\end{figure}

\begin{figure}
\includegraphics[width=1\textwidth]{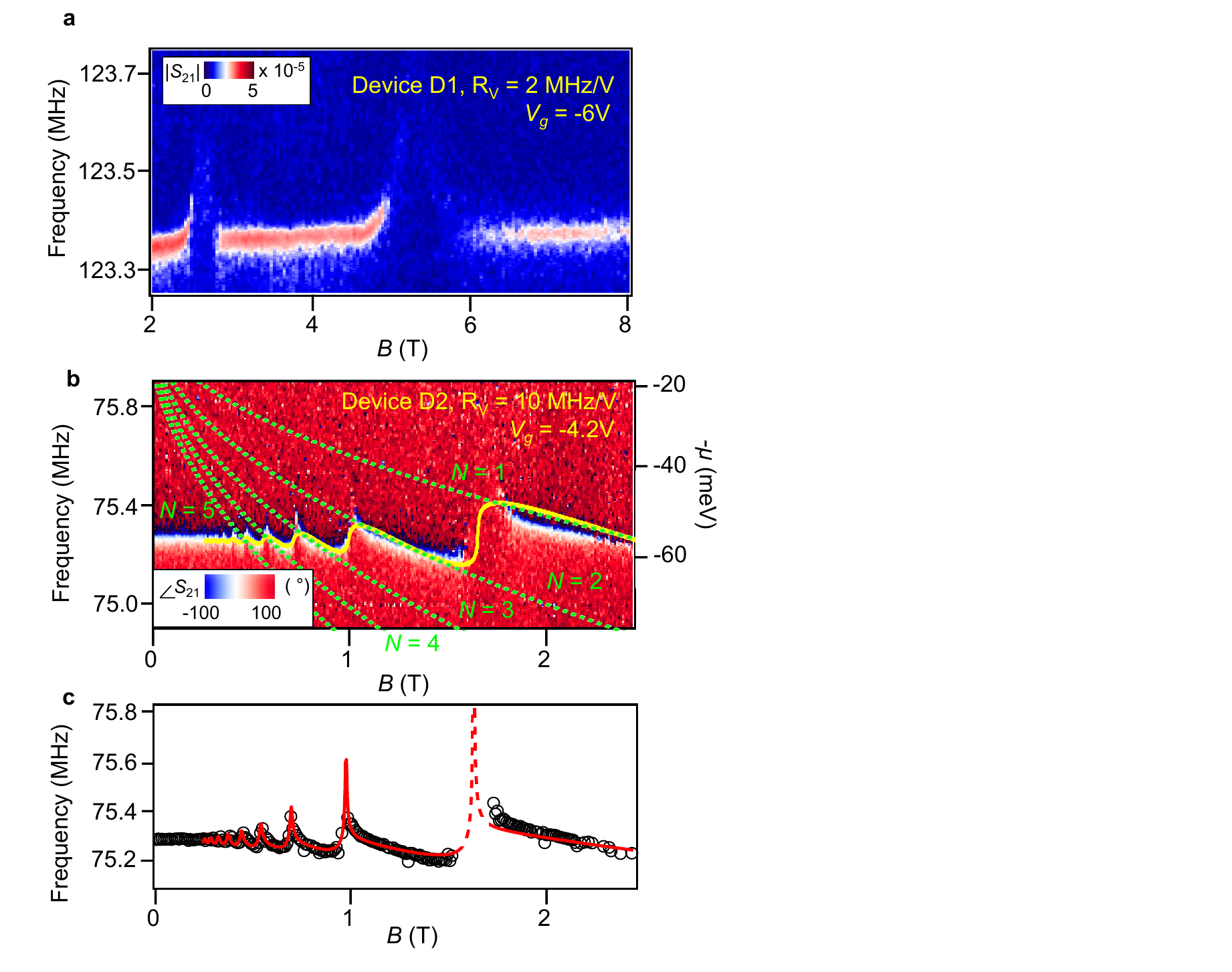}
\caption{Chemical potential variation induced frequency shifts. \textbf{(a)} Measured magnitude of $S_{21}$ transmission as function of applied magnetic field for device D1. Due to the small frequency tunability, there is no obvious mechanical resonance shift within
single LL. . Test conditions: $T$ = 4.3 K, $V_g$ = -6 V, drive power is -62 dBm. \textbf{(b)} Similar measurement for device D2, which has larger frequency tunability. The corresponding chemical potential variation is overlaid in yellow. The green dotted lines show the LL energies for N = 1 to 5. Test conditions: $T$ = 4.3 K, $V_g$ = -4.2 V, drive power is -68 dBm. \textbf{(c)} Complete fitted result (red curve) to the data shown in \textbf{(b)}. The dashed line indicates where the model is not expected to be accurate.}
\end{figure}

\begin{figure}
\includegraphics[width=1\textwidth]{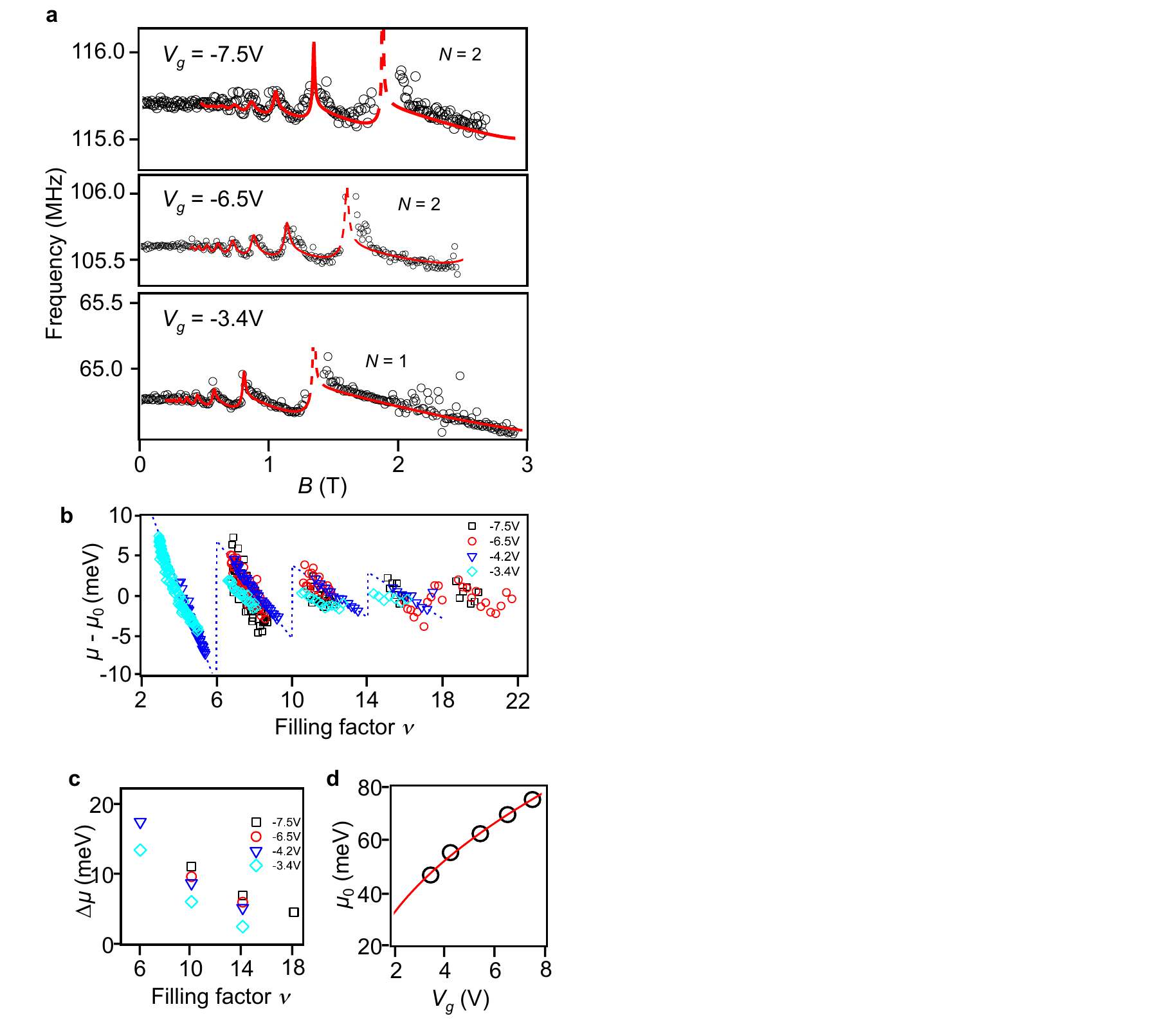}
\caption{Chemical potential evolutions and overall fits to experimental data. \textbf{(a)} Mechanical resonant frequency as functions for applied magnetic fields at different $V_g$, and corresponding fits (red curves). The dashed line for $V_g$ = -3.4 V indicates where the model is not expected to be accurate (see text). \textbf{(b)} Extracted chemical potential changes for different $V_g$, as function of filling factors. The dashed blue line shows the linear extrapolation used to determine the energy gaps at integer filling factors. \textbf{(c)} Energy gaps at different integer filling factors for various $V_g$. \textbf{(d)} Chemical potential at zero magnetic field, $\mu_0$ at different $V_g$. Red solid curve is $\sqrt{V_g}$ fit.}
\end{figure}

\end{document}